# The effect of hyperfine splitting on Stark broadening for three blue-green Cu I lines in laser-induced plasma

Andrey M. Popov*, Nikolay I. Sushkov, Sergey M. Zaytsev, and Timur A. Labutin

Laser Chemistry Division, Department of Chemistry, Lomonosov Moscow State University,
Leninskie Gory 1-3, Moscow 119234, Russia
*E-mail: popov@laser.chem.msu.ru

Abstract. Stark effect is observed in many natural and artificial plasmas and is of great importance for diagnostic purposes. Since this effect alters profiles of spectral lines, it should be taken into account when assessing chemical composition of radiation sources, including stars. Copper is one of the elements which studies of stellar atmospheres deal with. To this end, UV and visible Cu lines are used. However, there is a lack of agreement between existing data on their Stark parameters. It is therefore of interest to obtain new experimental data on these lines and to compare them to previous results. In this work, we have estimated Stark widths and shifts for three blue-green lines at 5105.54, 5153.24, and 5218.20 Å (corresponding transitions are $[3d^{10}4p]\ ^2P° \rightarrow [3d^94s^2]\ ^2D$ and $[3d^{10}4d]\ ^2D \rightarrow [3d^{10}4p]\ ^2P°$) observed in a "long-spark" laser-induced plasma. For the first time, we have accurately estimated an impact of hyperfine splitting on the profile shapes of the studied lines taking also into account the isotope shifts. We have shown that both effects considerably influence shift and width of Cu I line at 5105.54 Å, and shifts of Cu I lines at 5153.24 and 5218.20 Å.

## 1. INTRODUCTION

Photosphere is a part of stellar atmosphere where visible emission predominantly originates. If temperature of this layer exceeds $10^4$ K, hydrogen and other elements get considerably ionized. This is the case with the hottest stars of spectral classes O, B, A0 and white dwarfs (de Andrés-García,Colón & Fernández-Martínez 2018; Hamdi et al. 2018; Dimitrijevic 1989). Stark broadening dominates the line profiles under such conditions. Since the Stark effect grows stronger the greater the principal quantum number of a transition is, it may be significant even at very low electron number densities in interstellar nebulae (Hentschel 2009; Dimitrijevic 1989).

Elemental stellar abundances are determined using pressure-broadened profiles of absorption lines (Hinkel et al. 2016), which can be considerably disturbed due to the Stark effect. Therefore, it should be taken into account for a correct evaluation of a star's composition. Besides this, Stark broadening may be used for electron density determination even in non-equilibrium plasma source (Konjević 1999).

However, Stark widths and shifts are well known only for a small number of spectral lines compared to their total known amount. Indeed, the widely used atomic spectral line databases (Kramida et al. 2017; Kurucz & Bell 1995) contain several hundred thousand or millions of lines, while Stark parameters are known (both calculated and experimental ones) only for a few thousand transitions (Konjevic & Roberts 1976; Konjević & Wiese 1976; Konjević,Dimitrijević & Wiese 1984a; b; Konjević & Wiese 1990; Konjević et al. 2002; Lesage 2009; Sahal-Bréchot,Dimitrijević & Moreau 2017b). Unfortunately, calculated values often differ between themselves due to different methods used and do not agree with experimental results, especially in the case of heavy elements with a complex electronic structure. This situation leads to practical difficulties, because the parameters are needed for spectral modelling (Short,Jason & Lindsey 2018), including cutting-edge non-LTE modelling of stellar atmospheres (Andrievsky et al. 2018; Yan et al. 2016; Yan,Shi & Zhao 2015), determination of elemental abundances in heavenly bodies (Hinkel et al. 2016), as well as for estimation of radiative transfer through stellar atmospheres and subphotospheric layers (Tagirov,Shapiro & Schmutz 2017), opacity calculations (Krief,Feigel & Gazit 2016) and other astrophysical topics. Therefore, experimental determination of Stark effect parameters is a highly relevant topic both for astrophysics as well as for diagnostics of laboratory plasmas.

Since it is hardly possible to study all the necessary parameters within a single work, we focused on selected copper lines for the following reasons. Copper is present in stellar atmospheres in a considerable amount and is usually involved in studies of stellar metallicities (Mishenina et al. 2011; Cowley et al. 2016), classification of stars and evolution of the Galaxy (Yan et al. 2015). The astrophysical site for the synthesis of copper is not yet well established (Shi et al. 2014), and future research is required. Non-LTE effects, especially in metal-poor stars, may considerably distort results of copper abundance calculations and classification of stars (Yan et al. 2016). The strongest blue-green Cu I lines (5105.54, 5153.24 and 5218.20 Å) belonging to two transitions, i.e. $^2P° \to {}^2D$ and $^2D \to {}^2P°$, lie in one of the transparency windows of the Earth's atmosphere. After UV resonant transitions at 324 and 327 nm, whose upper levels can undergo strong departure from LTE due to depopulation caused by photoionization (Shi et al. 2014), these lines are among the strongest copper transitions, this is why they are intensive enough for astrophysical studies (Yan et al. 2016; Zhao et al. 2016) even in spectra of metal-poor stars. Moreover, broadening and shift for the $^2D \to {}^2P°$ multiplet are expected to be two orders of magnitude greater than those for the resonant doublet (Sahal-Bréchot et al. 2017b), which would be advantageous for plasma diagnostics.

Experimental Stark parameters for Cu lines in UV and blue spectral regions were recently published (Skočić et al. 2013; Burger et al. 2014). However, existing data for lines around 515 nm have discrepancies between each other. For example, the measured Stark widths for Cu I 5153.24 Å range from 190 to 2565 pm at $N_e = 10^{17}$ cm$^{-3}$ ((Fleurier 1986) vs. (Ovechkin & Sandrigailo 1969)), while calculation through various assumptions gives values from 63 to 576 pm at $N_e = 10^{17}$ cm$^{-3}$ ((Konjević & Konjević 1986) vs. (Pichler 1972)). Moreover, there is no information how hyper-fine splitting and isotope shifts influence the observed Stark parameters of Cu I lines. Obviously, new, more accurate experimental data on the Stark parameters of blue-green copper lines are needed to reduce the uncertainties.

Laser-induced plasmas (LIPs) are quite extensively used for the determination of Stark parameters (Konjević et al. 2002; Skočić et al. 2013; Popov et al. 2017), relatively easy to work with, have considerable electron number density and allow selecting different working conditions merely by changing temporal parameters of signal acquisition. Generating of long plasma ("long spark") instead of spherical plasma can provide the following advantages for Stark parameter measurements: (i) illumination of the whole available focal plane of spectrograph, thus increasing the vertical dimension of plasma image on the detector 5–7-fold, which leads to an enhancement of signal-to-noise ratio; (ii) long spark is more homogeneous than spherical one, as we demonstrated earlier (Popov et al. 2016; Popov et al. 2017), resulting in lesser optical thickness and lower experimental errors (<10%).

## 2. EXPERIMENTAL

*2.1. Set-up description.*

The long laser plasma was observed in the side-on mode using the optical configuration described earlier (Popov et al. 2016). Radiation of the second harmonics (532 nm, beam diameter = 6 mm, 8 ns, 82 mJ/pulse) of the Q-switched Nd:YAG laser (model LS-2134UTF, LOTIS TII, Belarus) was focused on the surface of magnesium alloy by a glass cylindrical lens ($f$ = 500 mm). It produced 14 mm long plasma on the surface. Short focal length aspheric quartz lens ($f$ = 30 mm) was placed ~6 cm above the sample surface to provide maximal collection of plasma radiation with minimal aberrations on the end of an optical fiber with the reduction ratio of 2:1. The fiber projected an image of the long plasma onto the slit (25 μm) of Czerny Turner spectrograph (HR-320, ISA, USA) providing high resolution with 3600 lines/mm grating. Plasma emission spectra were recorded by high-speed ICCD camera (Nanogeit-2V, NANOSCAN, Russia) with the use of laboratory-made software previously described elsewhere (Zaytsev et al. 2014). Since plasma optical thickness leads to the overestimation of Stark widths (Konjević 1999), targets with low copper content have to be used. We used an Al-Cu-Li alloy (4% Cu) doped with manganese (~100 ppm) and magnesium (0.89%) as a target.

## 2.2. Spectrograph calibration

Three copper lines in the blue-green region of spectrum were studied, namely, at 5105.5 ($[3d^{10}4p]$ $^2P_{3/2}° \rightarrow [3d^94s^2]$ $^2D_{5/2}$), 5153.2 ($4d$ $^2D_{3/2} \rightarrow 4p$ $^2P_{1/2}°$), and 5218.2 Å ($4d$ $^2D_{5/2} \rightarrow 4p$ $^2P_{3/2}°$). To estimate plasma parameters we used the ranges with central wavelengths at 2800 Å, 4030 Å and 5170 Å. Closeness of the studied lines to the grating cutoff wavelength provided an extremely high resolution (up to 200,000 at 5200 Å), but the reciprocal linear dispersion changed significantly within the working range. To precisely measure spectral shifts, we build wavelength calibration function of the spectrograph as precisely as possible (**Figure 1**). For these purposes, we used spectra of the Al alloy, pure iron, copper, titanium, and concrete at longer delays as possible (10 μs) to avoid significant Stark shifts of the lines. As one can see, calibration functions significantly deviate from linear one, therefore, we have used a quadratic function (e.g. $\lambda=a+b\times n+c\times n^2$) for these purposes. Estimation of wavelength uncertainty Δλ in the center of spectral window was obtained from uncertainties of calibration function coefficients. Instrumental width was estimated as 4.4 pm (5120 Å range), 3.8 pm (5153.2 Å range), 2.5 pm (5218.2 Å range), and 26.3 pm (2800 Å range). These values were subtracted from the experimental line widths.

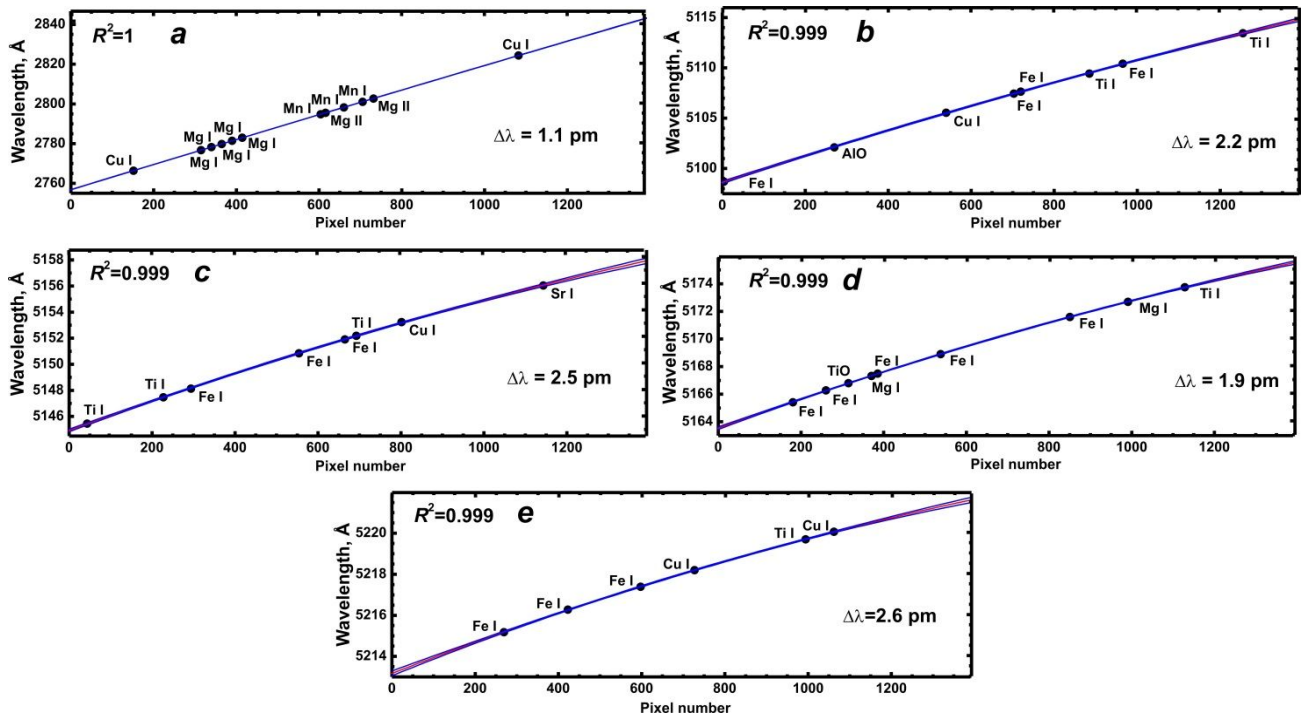

**Figure 1.** Spectrograph wavelength calibration curves in the working ranges. Lines used for calibration are shown by points, and emitting particles are indicated nearby. Red lines indicate the 95% confidence interval, and blue lines — the 95% prediction interval. Δλ stands for the wavelength uncertainty (width of the 95% prediction interval).

*2.3. Temporal acquisition parameters*

The position and width of an emission line can be altered by electron density variations within gate time resulting in the apparent asymmetry of lines. This might be overcome by using the shortest possible gate for signal registration. Thus, the compromise between this factor and decrease of signal-to-noise ratio (SNR) with short exposure time was that SNR did not fall below 25 and the ratio of delay to gate was not < 10 as recommended by Aragon and Aguilera (Aragón & Aguilera 2008). At each delay, 5 spectra were accumulated for 40 laser shots at 5 different spots, thus yielding 25 spectra. Temporal acquisition parameters are given in **Table 1**. Since observed profile wings of two Cu I lines (5153.2 Å and 5218.2 Å) were out the spectral window of the spectrograph at 500 ns delay, we omitted it for these lines.

Table 1. Temporal acquisition parameters (delay/gate, μs) for the studied lines.

| Al II 2816.19 Å | Mg I 5172.68 Å | Mn I 4030-4050 Å | Cu I 5105.54 Å | Cu I 5153.23 Å | Cu I 5218.20 Å |
|---|---|---|---|---|---|
| 0.5/0.02 | 0.5/0.05 | 0.5/0.2 | 0.5/0.05 | – | – |
| 0.75/0.03 | 0.75/0.07 | 0.75/0.2 | 0.75/0.08 | 0.75/0.05 | 0.75/0.05 |
| 1/0.05 | 1/0.1 | 1/0.5 | 1/0.1 | 1/0.1 | 1/0.1 |
| 1.5/0.1 | 1.5/0.15 | 1.5/0.5 | 1.5/0.2 | 1.5/0.1 | 1.5/0.1 |
| 3/0.3 | 3/0.2 | 3/1 | 3/0.5 | 3/0.15 | 3/0.15 |

*2.4. Plasma diagnostics*

Plasma temperature was estimated using the Boltzmann two-lines method for neutral manganese lines at 4030.76 Å and 4041.36 Å (their parameters were retrieved from NIST Database (Kramida et al. 2017)), since they provide an appropriate energy gap between the upper states (2.1 eV), are close to each other and are not likely self-absorbed due to the low content of manganese (see Section 2.1). Plasma temperature $T$ and its uncertainty $\Delta T$ was obtained from the equations for the two-lines method (Labutin et al. 2013):

$$T = \frac{|E_2 - E_1|}{k}\left(\ln\frac{A_2 g_2 I_1 \lambda_1}{A_1 g_1 I_2 \lambda_2}\right)^{-1}, \tag{1}$$

$$\frac{\Delta T}{T} = \left|\frac{kT}{E_2 - E_1}\right|\sqrt{\left(\frac{\Delta I_1}{I_1}\right)^2 + \left(\frac{\Delta I_2}{I_2}\right)^2 + \left(\frac{\Delta A_1}{A_1}\right)^2 + \left(\frac{\Delta A_2}{A_2}\right)^2}, \tag{2}$$

where *E*, *A*, *g*, *I* and *λ* are the energy of the upper level, Einstein's coefficient, degeneracy of the upper level, line intensity (area of Lorentzian profile) and its wavelength, respectively.

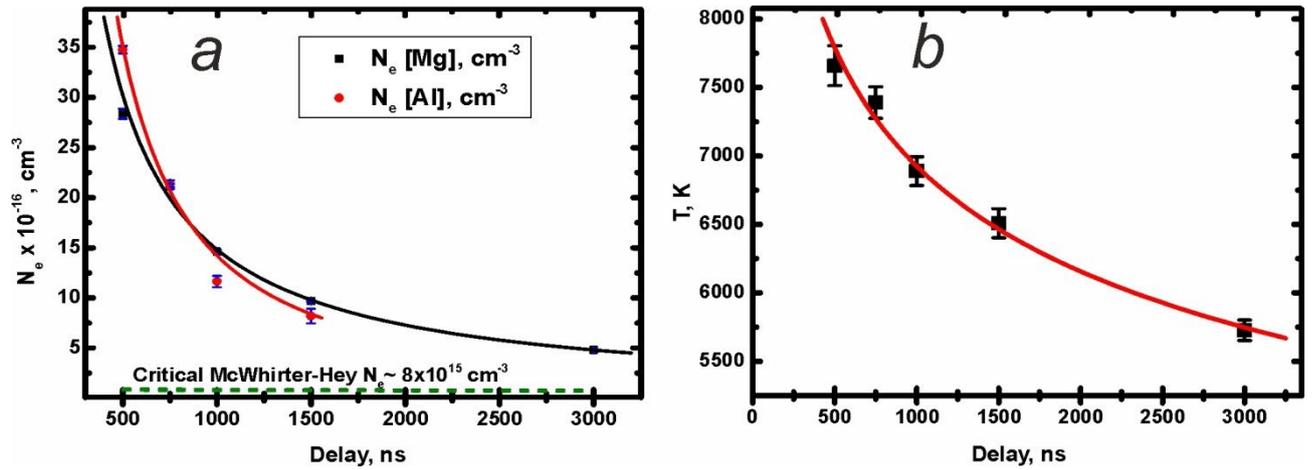

**Figure 2.** Evolution of electron density (*a*) calculated by Stark broadening for Mg I 5172.68 Å line (black points and curve) and Al II 2816.19 Å line (red points and curve) and plasma temperature (*b*). For comparison, the critical value of $N_e$ calculated according to the McWhirter-Hey criterion for the existence of LTE is marked by dotted line.

Electron number density $N_e$ (cm$^{-3}$) was estimated by Stark broadening of Al II 2816.19 Å and Mg I 5172.68 Å lines. Since there is no significant Debye shielding or ion broadening in these cases, a simplified expression for Stark width may be used (Konjević 1999):

$$FWHM_S(N_e, T) = 2w_e(T)10^{-16}N_e \quad, \qquad (3)$$

where $FWHM_S$ is Stark full width at half-maximum and $w_e$ is the electron impact parameter, taken from (Griem 1974) for Mg I line and from (Colón et al. 1993) for Al II line with interpolation to match the actual plasma temperature. Electron densities, calculated from Mg and Al line broadening, are in good agreement (**Figure 2**). Observed deviation may be caused by Al II and Mg I emission coming from different zones of the plasma having different parameters. The Al II line is weak, and its intensity is rapidly decreasing with time, so it was not measured at the latest delay.

To check if local thermodynamic equilibrium (LTE) can exist in the plasma, we calculated the critical $N_e$ as devised by McWhirter and Hey (Cristoforetti et al. 2010):

$$N_e > \frac{2.55 \times 10^{11}}{\langle \bar{g} \rangle} T^{\frac{1}{2}} (\Delta E)^3 , \qquad (4)$$

where $T$ is in K, $\langle \bar{g} \rangle$ is a dimensionless averaged Gaunt factor, $\Delta E$ (in eV) is the energy gap between upper and lower levels. For neutral atoms and singly charged ions Van Regemorter has proposed (van Regemorter 1962) the estimation of averaged Gaunt factor, which depends on the dimensionless ratio $\Delta E/kT$. The minimal value of $N_e$ is $9.0 \times 10^{15}$ cm$^{-3}$ for Al II 2816.19 Å and $7.2 \times 10^{15}$ cm$^{-3}$ for Mg I 5172.68 Å. Thus, the measured $N_e$ values are well above these limits. Trends of electron density and temperature are symbate (**Figure 2**).

Since the main obstacle for accurate estimation of plasma parameters is self-absorption of emission lines (Konjević, Ivković & Jovićević 2010), we have evaluated the absorption coefficients of the line used for diagnostics under the conditions of the long laser spark. The experimental data was fitted by synthetic spectra generated in approximation of static homogeneous plasma under LTE condition, accounting line broadening and absorption, more detailed description can be found in (Zaytsev, Popov & Labutin 2019). We supposed the equality of the elemental composition of the target and plasma, and the experimentally determined values of $T$ and $N_e$ are used (*see* **Figure 2**) to obtain estimates of absorption coefficients. The resulting values are collected in **Table 2**. One can see that the self-absorption of Al and Mn lines used for the determination of $N_e$ and $T$ is negligible, and a slight influence of the self-absorption can be observed at the latest delay for the Mg line. The closeness of the observed intensities within Mn and Mg multiplets to theoretical values based on *LS*-coupling also proves the absence of self-absorption.

**Table 2. Absorption coefficients for the studied lines.**

| Delay, μs | Al II 2816.19 Å | Mg I 5172.68 Å | Mn I 4030.76 Å | Mn I 4041.36 Å |
|---|---|---|---|---|
| 0.5 | <0.01 | 0.06 | 0.02 | <0.01 |
| 0.75 | <0.01 | 0.07 | 0.02 | <0.01 |
| 1 | <0.01 | 0.07 | 0.02 | <0.01 |
| 1.5 | <0.01 | 0.08 | 0.02 | <0.01 |
| 3 | <0.01 | 0.10 | 0.04 | <0.01 |

### 3. RESULTS AND DISCUSSION

*3.1 Estimation of line widths due to different broadening mechanisms*

We have considered the influence of Doppler, pressure, resonance, and Van der Waals broadening mechanisms on the widths of Cu I lines. Calculation methodology is described elsewhere (Popov et al. 2016; Konjević 1999); **Table 3** lists widths calculated for Cu lines, and broadening of Mg and Al lines due to these effects may be considered negligible (Popov et al. 2016). Generally, the widths listed in **Table 3** were an order or two of magnitude lower than observed ones (up to hundreds of pm, *see also* Section 3.3), that we could neglect the influence of above mentioned mechanisms with the exception of Doppler broadening for copper line at 5105.54 Å at long delays.

**Table 3. Cu line widths due to different broadening mechanisms (at 1 μs delay, $T$=6900 K)**

| Broadening mechanism | Width, pm | | |
|---|---|---|---|
| | Cu 5105.54 Å | Cu 5153.24 Å | Cu 5218.20 Å |
| **Doppler** | 3.8 | 3.8 | 3.9 |
| **Pressure** | 0.17 | 0.17 | 0.17 |
| **Resonance** | 1.0 | 0.9 | 1.6 |
| **Van der Waals** | 0.53 | 0.18 | 0.18 |

*3.2. Hyperfine structure of the studied lines*

Natural copper consists of two odd isotopes, namely $^{63}$Cu and $^{65}$Cu, with the ratio of 0.6917:0.3083. Both of them have nuclear spin, $I=3/2$ (Radzig & Smirnov 1985), resulting in a large nuclear magnetic dipole ($\mu(^{63}$Cu$) = 2.22$ and $\mu(^{65}$Cu$) = 2.38$ (Radzig & Smirnov 1985)) and small electric-quadrupole moments. The total angular momentum $F$ of an atomic state is then given by

$$|J - I| \leq F \leq |J + I|, \quad (5)$$

where $J$ is the total angular momentum of electrons. According to Eq. (6), five levels of the lines under investigation split into four *hfs*-components, while the $4p$ $^2$P$_{1/2}°$ state splits into two components only. In order to predict the observed *hfs*-components the appropriate selection rules need to be applied, which, in the absence of an external magnetic field, are given as $\Delta F = 0, \pm 1$ and $F=0 \leftrightarrow F=0$ is forbidden (Fischer, Hühnermann & Kollath 1967). Therefore, the allowed *hfs*-transitions are illustrated in **Figure 3** by the vertical downwards arrows. Two lines, 5105.54 Å and 5218.20 Å, have nine *hfs*-components, but their total number reaches 18 taking into account the $^{65}$Cu isotope. The 5153.24 Å line has six *hfs*-components (or 12 lines with the $^{65}$Cu isotope).

The frequency shifts, $\Delta v$, of the *hfs*-components relative to the centers of gravity of the levels can be calculated from (Kopfermann & Schneider 1958):

$$\Delta v = \frac{AC}{2} + \frac{B\frac{3}{2}C(C+1) - 2I(I+1)J(J+1)}{4 \quad IJ(2I-1)(2J-1)}, \quad (6)$$

where $A$, $B$ and $C$ are the magnetic dipole, electric quadrupole splitting factors, respectively, and $C = F(F+1) - J(J+1) - I(I+1)$. The quadrupole interaction constant $B$ is zero for states where $F \leq 1/2$ owing to the spherical symmetry of the charge distribution. The available values of $A$ and *B* for $^{63}$Cu lines are presented in **Table 4**. To our best knowledge, there are data on *A* and *B* constants only for three states in $^{65}$Cu isotope: $^2D_{5/2}$ (Fischer et al. 1967), $^2P_{1/2}°$ (Fischer, Hühnermann & Kollath 1966; Bergström, Peng & Persson 1989) and $^2P_{3/2}°$ (Bergström et al. 1989), since the known isotope shifts for each line were used to calculate the positions of *hfs*-components of $^{65}$Cu isotope. In this work, we have calculated shifts and relative intensities of *hfs*-components (**Table 5**) for both isotopes to estimate the effect of *hfs*-splitting on the observed line profile. Relative intensities of *hfs*-components were calculated with the use of quantum numbers *I*, *J*, and *F* (*S*, *L*, and *J*, respectively, in original work) in accordance with transition-specific multiplet-normalized line strengths given in (Axner et al. 2004). Frequency shifts for the first transition (5105.5 Å) are in a good agreement with observations of (Tenenbaum et al. 1980).

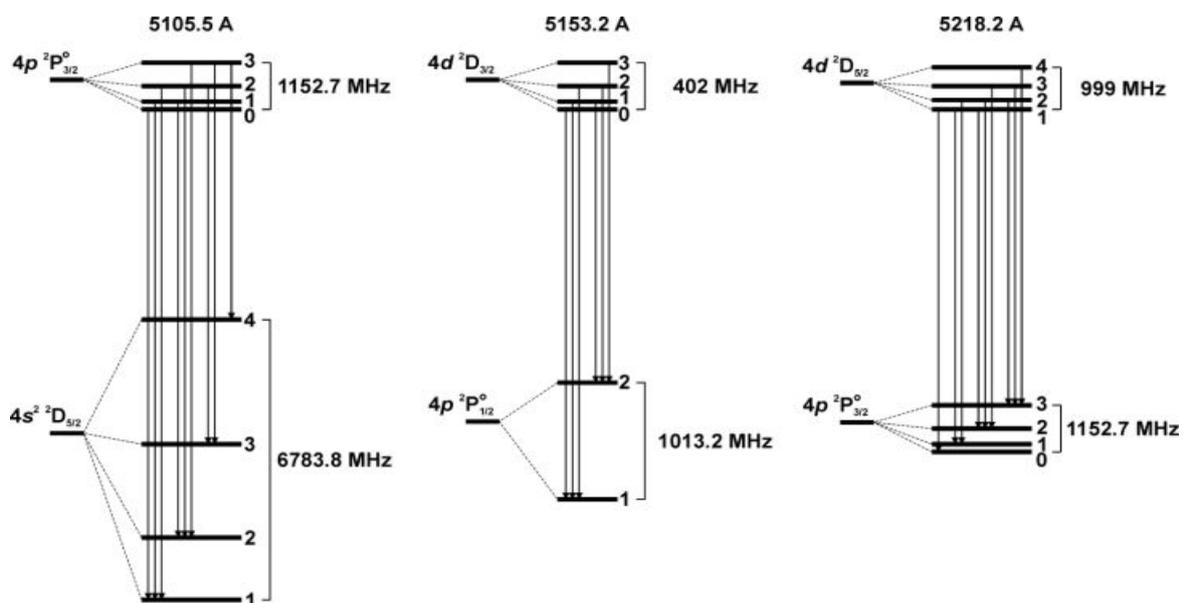

**Figure 3.** A scheme illustrating *hfs*-splitting and the allowed *hfs*-transitions within three $^{63}$Cu I lines under investigation: 5105.5 Å, 5153.2 Å and 5218.2 Å. $F$ values of split levels are indicated on the right, while terms of initial levels are on the left.

**Table 4.** *hfs*-parameters of $^{63}$Cu atomic levels used to calculate the hyperfine structure

| Configuration | Term | Energy, cm$^{-1}$ | A, MHz | B, MHz | Reference |
|---|---|---|---|---|---|
| $3d^94s^2$ | $^2D_{5/2}$ | 11202.618 | 749.2 | 187.1 | (Fischer et al. 1967) |
| $3d^{10}4p$ | $^2P_{1/2}°$ | 30535.324 | 506.6 | 0.0 | (Bergström et al. 1989) |
| $3d^{10}4p$ | $^2P_{3/2}°$ | 30783.697 | 194.3 | -26.2 | (Bergström et al. 1989) |
| $3d^{10}4d$ | $^2D_{3/2}$ | 49935.195 | 67.0 | 0.0 | (Ankush & Deo 2014) |
| $3d^{10}4d$ | $^2D_{5/2}$ | 49942.051 | 111.0 | 0.0 | (Ankush & Deo 2014) |

The *hfs*-components of the Cu I lines for both isotopes, $^{63}$Cu and $^{65}$Cu, and the resulting profiles for the Cu I lines as superposition of Voigt profiles calculated for each *hfs*-component are presented in **Figure 4**. Lorentzian part of Voigt profile was a sum of Stark width ($w_S$) and instrumental broadening ($w_I$), while Gaussian part was Doppler broadening ($w_D$) calculated for our temperatures. Stark parameters (width $w_S$ and shift $d_S$) for particular temperature were calculated following equations given in (Sahal-Bréchot, Dimitrijević & Ben Nessib 2017a):

$$\lg w_S = a_0 + a_1 \lg T + a_2 (\lg T)^2 \text{ and } d_S/w_S = b_0 + b_1 \lg T + b_2 (\lg T)^2, \qquad (9)$$

where coefficients $a_{0-2}$ and $b_{0-2}$ were retrieved from the STARK-B database (Sahal-Bréchot et al. 2017b). Initially we have neglected Stark and Doppler effects and considered the *hfs* and instrumental function only. As one can see, the *hfs* width for Cu I 5105.5 Å as well as its isotope shift are larger than in the case of other copper lines, and, moreover, they are comparable with those resulting from broadening processes (see **Figure 4, a**). Therefore, the profile of the Cu I line has several peaks, unlike resonance Cu I lines (Burger et al. 2014) and other copper lines under investigation (see **Figure 4, b** and **c**) which have profiles close to Lorentzian one due to instrumental broadening. Convolution of the *hfs*-patterns with the profile accounting all broadening mechanisms under our experimental conditions ($N_e$, *T*) provided "apparent" Stark widths and shifts. Although the line at 5105.54 Å has a pronounced hyperfine structure, it also adequately described by the Lorentzian function. "Apparent" Stark parameters produced from theoretical ones (Babina et al. 2003, Konjević & Konjević 1986: eq.2 for multiplet $4d\ ^2D→4p\ ^2P$, Zmerli et al. 2010b) resulting from linear approximation ($d = d_0 + d_S×N_e$ and $FWHM$-$w_I = w_0 + w_S×N_e$) are given in **Table 6.** Non-zero value of $w_0$ for Cu I 5105.5 Å line is a result of *hfs*-width of ~15 pm, since it is comparable with Stark width and larger than Doppler width under our conditions (**Figure 4, a**). Other Cu I lines were free of such an effect due to relatively large $w_S$. Therefore, *hfs*-width should be taken into account when considering experimental data on the Cu I 5105.5 line. In the meanwhile, hyper fine splitting is a reason of non-zero shift of a line center $d_0$ for the copper lines. Since the parameters for copper lines at 5153.2 Å and 5218.2 Å, given in STARK-B Database (Sahal-Bréchot et al. 2017b), are the same, behavior of their dependencies of "apparent" Stark parameters on $N_e$ coincided.

**Table 5.** Frequency shifts (Δν, in MHz), wavelength shifts (Δλ, in pm), line strengths (*f*) and relative intensities (*I*) of *hfs*-components of Cu I lines under investigation with respect to isotope shifts (IS, in MHz). The centers of gravity of transitions (in Å) are given for $^{63}$Cu isotope.

| Transition | | | | $^{63}$Cu | | | $^{65}$Cu | | |
|---|---|---|---|---|---|---|---|---|---|
| Terms, wavelength and IS | $F_{upper}$ | $F_{lower}$ | *f* | Δν | Δλ | *I* | Δν | Δλ | *I* |
| $4p\ ^2P_{3/2}°→4s^2\ ^2D_{5/2}$ 5105.5 Å IS=2128 (Gerstenberger, Latush & Collins 1979) | 3 | 4 | 0.3750 | -2309.4 | 6.02 | 100.0 | -181.4 | 0.47 | 44.9 |
| | 2 | 3 | 0.2333 | 129.3 | -0.34 | 62.2 | 2257.8 | -5.89 | 27.9 |
| | 3 | 3 | 0.0583 | 705.6 | -1.84 | 15.6 | 2834.1 | -7.39 | 7.0 |
| | 1 | 2 | 0.1312 | 2006.3 | -5.23 | 35.0 | 4135.2 | -10.78 | 15.7 |
| | 2 | 2 | 0.0729 | 2390.5 | -6.23 | 19.4 | 4519.4 | -11.79 | 8.7 |
| | 3 | 2 | 0.0042 | 2966.9 | -7.74 | 1.1 | 5095.7 | -13.29 | 0.5 |
| | 0 | 1 | 0.0625 | 3321.7 | -8.66 | 16.7 | 5450.8 | -14.22 | 7.5 |
| | 1 | 1 | 0.0563 | 3513.8 | -9.16 | 15.0 | 5642.9 | -14.72 | 6.7 |
| | 2 | 1 | 0.0063 | 3898.0 | -10.17 | 1.7 | 6027.1 | -15.72 | 0.8 |
| | - | - | Σ=1 | - | - | - | - | - | - |
| $4d\ ^2D_{3/2}→4p\ ^2P_{3/2}°$ 5153.2 Å IS=-363 (Elbel & Fischer 1961) | 1 | 2 | 0.0312 | -564.2 | 1.50 | 7.1 | -927.2 | 2.46 | 3.2 |
| | 2 | 2 | 0.1563 | -430.2 | 1.14 | 35.7 | -793.2 | 2.11 | 16.0 |
| | 3 | 2 | 0.4375 | -229.2 | 0.61 | 100.0 | -592.2 | 1.57 | 44.9 |
| | 0 | 1 | 0.0625 | 382.0 | -1.01 | 14.3 | 19.0 | -0.05 | 6.4 |
| | 1 | 1 | 0.1563 | 449.0 | -1.19 | 35.7 | 86.0 | -0.23 | 16.0 |
| | 2 | 1 | 0.1563 | 583.0 | -1.55 | 35.7 | 220.0 | -0.58 | 16.0 |
| | - | - | Σ=1 | - | - | - | - | - | - |
| $4d\ ^2D_{5/2}→4p\ ^2P_{3/2}°$ 5218.2 Å IS=-354 (Elbel & Fischer 1961) | 2 | 3 | 0.0042 | -805.3 | 2.19 | 1.1 | -1159.3 | 3.16 | 0.5 |
| | 3 | 3 | 0.0583 | -472.3 | 1.29 | 15.6 | -826.3 | 2.25 | 7.0 |
| | 1 | 2 | 0.0063 | -450.9 | 1.23 | 1.7 | -804.9 | 2.19 | 0.8 |
| | 2 | 2 | 0.0729 | -228.9 | 0.62 | 19.4 | -582.9 | 1.59 | 8.7 |
| | 1 | 1 | 0.0563 | -66.7 | 0.18 | 15.0 | -420.7 | 1.15 | 6.7 |
| | 4 | 3 | 0.3750 | -28.3 | 0.08 | 100.0 | -382.3 | 1.04 | 44.9 |
| | 3 | 2 | 0.2333 | 104.0 | -0.28 | 62.2 | -249.9 | 0.68 | 27.9 |
| | 1 | 0 | 0.0625 | 125.4 | -0.34 | 16.7 | -228.6 | 0.62 | 7.5 |
| | 2 | 1 | 0.1312 | 155.3 | -0.42 | 35.0 | -198.7 | 0.54 | 15.7 |
| | - | - | Σ=1 | - | - | - | - | - | - |

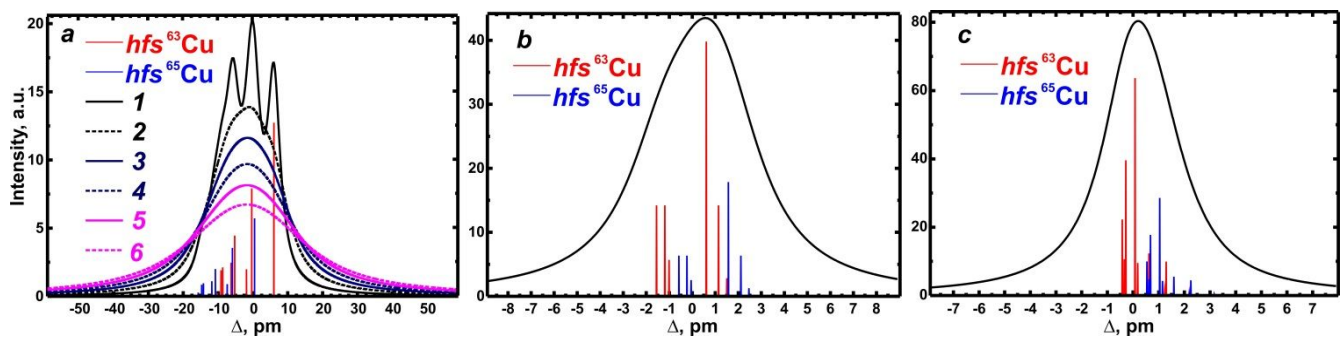

**Figure 4.** Spectral profiles of Cu I lines at 5105.5 Å (*a*), 5153.2 Å (*b*) and 5218.2 Å (*c*). *hfs*-components for $^{63}$Cu and $^{65}$Cu are shown in red and blue. Black solid curves (*1*) represent the profiles convoluted by the instrumental function only. Dashed and colored solid curves represent profiles accounting all broadening mechanisms under our experimental conditions: (*2*) $N_e = 5.05 \times 10^{16}$ cm$^{-3}$, $T = 5800$ K; (*3*) $N_e = 9.85 \times 10^{16}$ cm$^{-3}$, $T = 6500$ K; (*4*) $N_e = 1.55 \times 10^{17}$ cm$^{-3}$, $T = 6900$ K; (*5*) $N_e = 2.1 \times 10^{17}$ cm$^{-3}$, $T = 7400$ K; and (*6*) $N_e = 2.85 \times 10^{17}$ cm$^{-3}$, $T = 7700$ K.

**Table 6.** "Apparent" Stark parameters obtained for copper lines

|  | 5105.5 Å[†] | 5153.2 Å & 5218.2 Å[‡] |
|---|---|---|
| $w_S$, pm×cm$^3$ | $(8.1\pm0.2)\times10^{-17}/(7.7\pm0.3)\times10^{-17}$ | $(212.4\pm0.3)\times10^{-17}/(319.7\pm0.2)\times10^{-17}$ |
| $w_0$, pm | $10.5\pm0.4/9.2\pm0.4$ | $0.3\pm0.6/1.3\pm0.3$ |
| $d_S$, pm×cm$^3$ | $(9.95\pm0.05)\times10^{-17}/(8.55\pm0.05)\times10^{-17}$ | $-(37.1\pm0.8)\times10^{-17}/-(40.5\pm0.6)\times10^{-17}$ |
| $d_0$, pm | $-1.81\pm0.08/-2.03\pm0.09$ | $-(8\pm1)/-(7\pm1)$ |

[†] (Zmerli et al. 2010b)/(Babina et al. 2003)

[‡] (Zmerli et al. 2010b)/( Konjević & Konjević 1986, eq.2)

### 3.3. Measurements of Cu I Stark parameters

Intensities were normalized to unity to outline temporal evolution of profiles of the Cu I lines in **Figure 5**. Lorentzian approximation of the lines (Lorentzian part of the Voigt profile of 5105.54 Å line starting at 1000-ns delay) was used for determination of Stark parameters. Since the (4,5) band of the B$^2\Sigma^+ \rightarrow$ X$^2\Sigma^+$ system of AlO significantly disturbed the 5153.24 Å line profile in blue region (Hermann et al. 2015), we used only one side of the profile for fitting with Lorentzian function (**Figure 5,** *b*). We combined Lorentzian function with a linearly sloping background to fit the line at 5218.20 Å to diminish slight asymmetry (**Figure 5,** *c*) of the profile caused by the wing of strong Mg I 5183.6 Å.

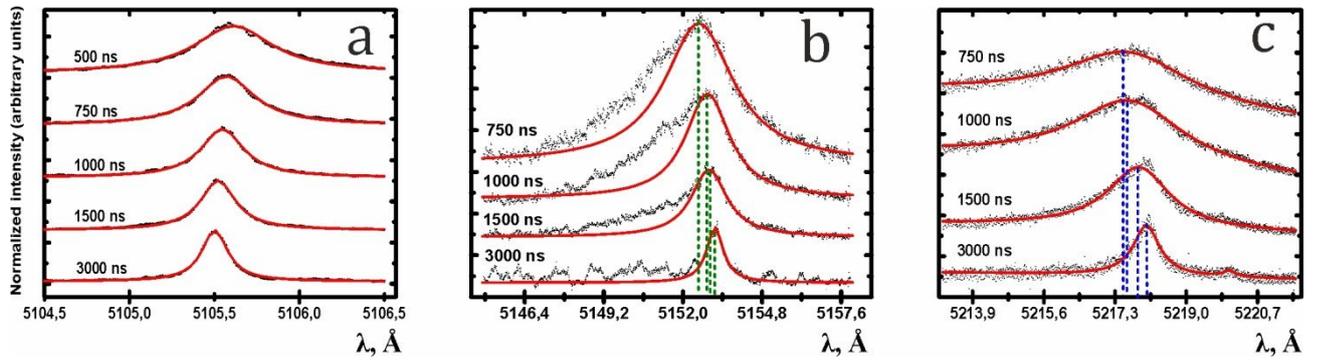

**Figure 5.** Temporal evolution of Cu I emission lines at 510.5 nm (*a*), 515.3 nm (*b*), and 521.8 nm (*c*). Black circles are experimental points; red curves are the best fit Lorentzian profiles. In the case of 515.3 nm line, red curves are cumulative profiles for multiline fit (Cu and interfering lines), and blue curves illustrate the shape of Cu line itself. Dashed lines indicate Stark-shifted central wavelengths of the lines.

Slopes of *FWHM* vs. $N_e$ and *d* vs. $N_e$ dependencies (**Figure 6**) served as Stark widths and shifts ($2w_e$ and $d_e$, respectively). Their values are $d_e$ = 4.6±1.2 pm and $2w_e$ = 17±0.7 pm for Cu I line at 5105.54 Å, $d_e$ = -32±2 pm and $2w_e$ = 152±9 pm for Cu I line at 5153.24 Å, and $d_e$ = -39±7 pm and $2w_e$ = 180±30 pm for Cu I line at 5218.20 Å. To compare obtained data with other experimental and theoretical values, we have taken the mean of temperature range as *T* (i.e. 6850 K). All values were scaled to $N_e = 10^{17}$ cm$^{-3}$. Stark parameters of the components of the same multiplet, namely the Cu lines at 5153.24 and 5218.20 Å, have to be close within measurement uncertainty (see Konjević, Ivković & Jovićević 2010). Therefore, our values are the same in this sense but parameters for Cu I line at 5153.24 Å are preferable to use because of small relative errors (about 6%). **Figure 7** provides a visual comparison of the resulting values of $2w_e$ and $d_e$ with known experimental and theoretical data, presented as a function of temperature if possible. We have also added the theoretical values (calculated "apparent" Stark parameters in **Table 6**) accounting the hyperfine splitting and isotope shift marked by a symbol "+hfs".

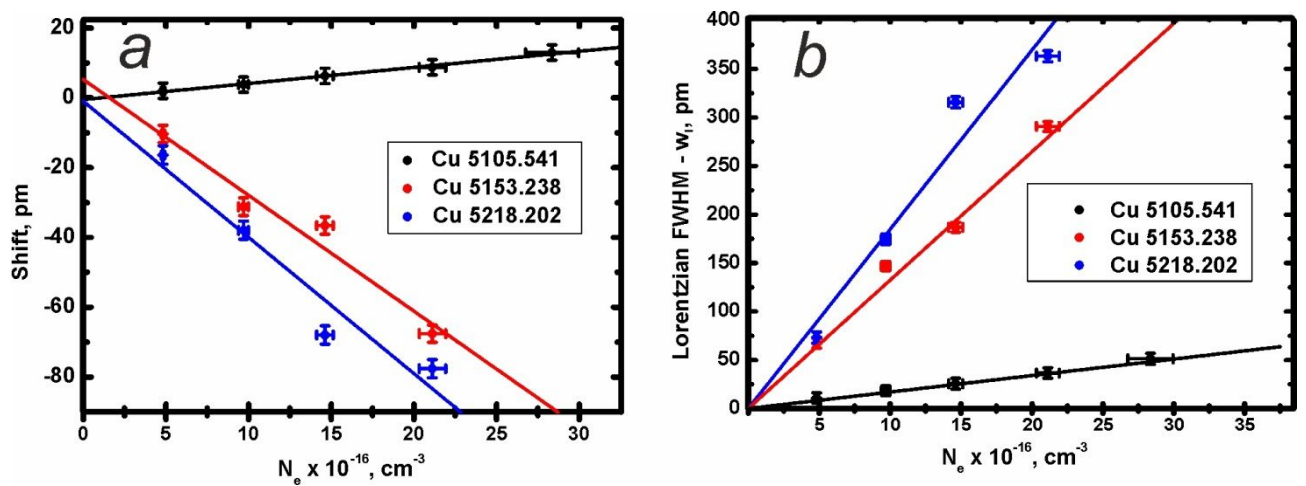

**Figure 6.** Stark shifts of Cu lines (*a*) and their FWHM (*b*) as a function of electron density. Experimental points with linear regression are shown.

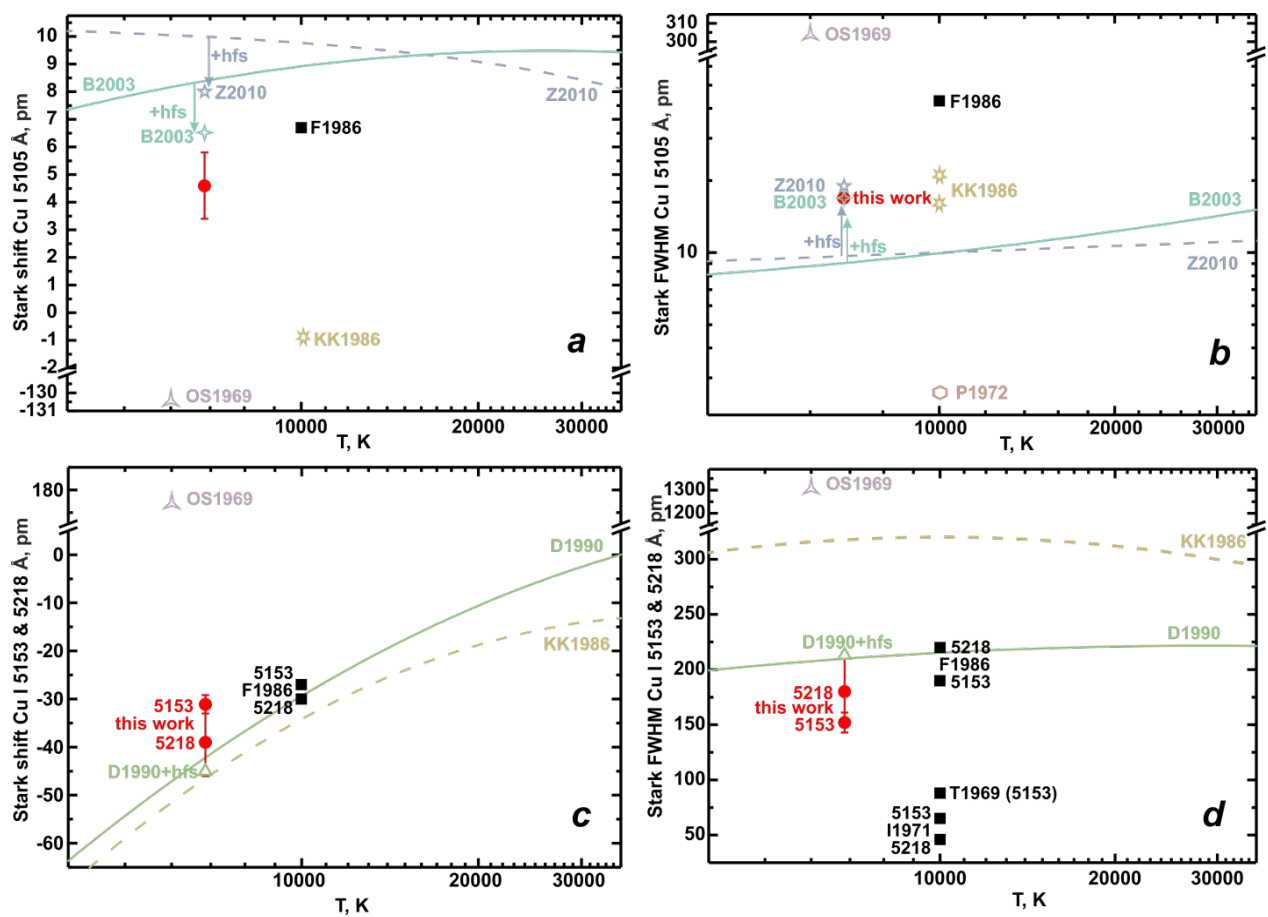

**Figure 7.** Experimental (filled points, this work in red) and theoretical (open points) Stark shifts (a,c) and Stark widths (b,d) for Cu I lines at 5105.54 Å (top), 5153.24 Å and 5218.20 Å (bottom) as a function of temperature. The curves are a result of fitting from B2003 (Babina et al. 2003), Z2010 (Zmerli et al. 2010b), KK1986 (Konjević & Konjević 1986, eq.2) and D1990 (Dimitrijević 1990). Data produced from theoretical curves by accounting of hyperfine structure are indicated as "+hfs". Other data are F1986 (Fleurier 1986), OS 1969 (Ovechkin & Sandrigailo 1969), P1972 (Pichler 1972), T1969 (Tonejc & Vujnović 1969), I1971 (Iwao & JayaRam 1971).

The Cu 5105.54 Å line is significantly less affected by Stark effect, which was already pointed out in (Song et al. 1999; Man et al. 2004). The only one experimental values of Stark parameters of this line (Fleurier 1986) has the obvious discrepancy with theoretical data on Stark FWHM. Fleurier reported that the 5105.54 Å line was very narrow so that instrumental broadening could affect it significantly. Simultaneously, our measurements are in agreement with "apparent" widths calculated for the values from (Babina et al. 2003) and (Zmerli et al. 2010b) as well as theoretical data from (Konjević & Konjević 1986). Although our value of Stark shift for the line is slightly lower than Fleurier's, Babina's and Zmerli's values, we should point out the calibration error for the spectral range, i.e. 2.5 pm (see Figure 1, c). In other words, the shift is the same in sense of uncertainty. Other available literature data (Ovechkin & Sandrigailo 1969, Pichler 1972) are strongly disagreed with our data and the values obtained by theory and Fleurier's experiment. It is probably due to the calculation from $C_4$-coefficients available in literature in 1970-th years. Another reason for such a discrepancy is very low electron number density in their experiment ($2.3 \times 10^{14}$ cm$^{-3}$), therefore, they did not measure Stark parameters at several different electron number densities, i.e. slope of $d(N_e)$ is unknown. Given that $N_e$ in their plasma is fairly low, direct extrapolation of $d$ to $N_e = 10^{17}$ cm$^{-3}$ is probably not reasonable. In any case, our values for Stark parameters are more preferable for the line due to high spectral resolution which was less than the observed width.

Although *hfs*-splitting and isotope shift are negligible for Stark width of Cu I lines at 5153.24 and 5218.20 Å, both effects are pronounced for Stark shifts of the lines (Figure 7, c and d). The observed differences between Stark parameters of both components of the multiplet $4d\ ^2D \rightarrow 4p\ ^2P°$ were 10% and 8% for shits and widths, respectively. Such a circumstance can be explained by the presence of third, small component of the multiplet near 5220 Å, i.e. Cu I line at 5220.07 Å visible for long delay time (**Figure 5, c**). Certainly the third component has presented over all times, but its width and shift as well as Stark parameters of stronger line at 5218.20 Å did not allow the accurate estimation of the parameters for the latter line. Therefore, we note the parameters obtained for the line 5153.24 Å are preferable for plasma diagnostics or other purposes due to smaller relative errors and the absence of strong interfering component. Self-absorption can not be an explanation of the observed differences because low content of copper (4%) and the short optical thickness of plasma along the direction of plasma view.

## 4. CONCLUSIONS

We have obtained experimental Stark widths and shifts for three neutral copper lines in the green region. These parameters are measured in long-spark laser-induced plasma obtained by vaporizing the Al—Cu—Li alloy, and are generally consistent with literature data. This shows that our experimental setup (a combination of the shortest possible gate with a long LIP) made it possible to minimize self-absorption, while the signal remained intensive enough to provide reliable results within a reasonable accumulation time. Time-resolved spectra acquisition enabled obtaining Stark parameters as slopes of relevant dependencies upon electron number density and calculating their uncertainties. Laser-induced plasma is a high-density source in which Stark effects are pronounced and relatively easy to measure. The role of the hyperfine structure components in the non-resonant Cu I line shapes formation was, for the first time, discussed taking also into account the isotope shift. It is shown that hyperfine structure and isotope shift additionally broad and shift profile of the Cu I line at 5105.54 Å. This should be taken into account in various calibration procedures based on these lines, especially in the case when measured electron densities are below $10^{18}$ cm$^{-3}$. Since there are considerable discrepancies between existing theoretical results, and experimental data are scarce, the presented results will be helpful for stellar studies, as well as for plasma diagnostics in general.

## 5. ACKNOWLEDGEMENTS

The reported study was supported by RFBR (grant No 18-33-01297) and the consideration of the influence of hyperfine structure on the observed Stark parameters of copper lines was supported by Russian Science Foundation (grant No 18-13-00269).

## 6. REFERENCES

Andrievsky S., Bonifacio P., Caffau E., Korotin S., Spite M., Spite F., Sbordone L., Zhukova A.V., 2018, MNRAS, 473, 3377
Ankush B.K., Deo M.N., 2014, J. Quant. Spectrosc. Radiat. Transfer, 134, 21
Aragón C., Aguilera J.A., 2008, Spectrochim. Acta Part B, 63, 893
Axner O., Gustafsson J., Omenetto N., Winefordner J.D., 2004, Spectrochim. Acta Part B, 59, 1
Babina E.M., Il'In G.G., Konovalova O.A., Salakhov M.K., Sarandaev E.V., 2003, in: Dimitrijevic M.S., Popovic L.C., Milovanovic N. eds., Iv Serbian Conference on Spectral Line Shapes. Astronomical Observatory, Belgrade, p. 163
Bergström H., Peng W.X., Persson A., 1989, Z. Physik D, 13, 203
Burger M., Skočić M., Nikolić Z., Bukvić S., Djeniže S., 2014, J. Quant. Spectrosc. Radiat. Transfer, 133, 589
Colón C., Hatem G., Verdugo E., Ruiz P., Campos J., 1993, J. Appl Phys., 73, 4752
Cowley C.R., Ayres T.R., Castelli F., Gulliver A.F., Monier R., Wahlgren G.M., 2016, ApJ, 826, 158
Cristoforetti G., De Giacomo A., Dell'Aglio M., Legnaioli S., Tognoni E., Palleschi V., Omenetto N., 2010, Spectrochim. Acta Part B, 65, 86
de Andrés-García I., Colón C., Fernández-Martínez F., 2018, MNRAS, 476, 793
Dimitrijevic M.S., 1989, Bull. Obs. Astron. Belgrade, 140, 111
Dimitrijević M.S., 1990, in: Veža D. ed., XV Summer school and International Symposium on the Physics of Ionized Gases. Institute of Physics of the University Zagreb, Dubrovnik, Yugoslavia,
Elbel M., Fischer W., 1961, Z. Physik, 165, 151
Fischer W., Hühnermann H., Kollath K.-J., 1966, Z. Physik, 194, 417
Fischer W., Hühnermann H., Kollath K.-J., 1967, Z. Physik, 200, 158
Fleurier C., 1986, in: Exton R.J. ed., Eighth International Conference on Spectral Line Shapes. A. Deepak Publishing, College of William and Mary, Williamsburg, VA, p. 67
Gerstenberger D.C., Latush E.L., Collins G.J., 1979, Optics Commun., 31, 28
Griem H., 1974, Academic Press, New York
Grishina N.A., Ilyin G.G., Salakhov M.K., Sarandaev E.V., 1998a, Coherent Optics and Optical Spectroscopy. Kazan, Russia, p. 43
Grishina N.A., Ilyin G.G., Salakhov M.K., Sarandaev E.V., 1998b, 19th SPIG Contributed papers. Zlatibor, Yugoslavia, p. 361
Hamdi R., Ben Nessib N., Sahal-Bréchot S., Dimitrijević M.S., 2018, MNRAS, 475, 800
Hentschel K., 2009, in: Greenberger D., Hentschel K., Weinert F. eds., Compendium of Quantum Physics. Springer Berlin Heidelberg, Berlin, Heidelberg, p. 738
Hermann J., Lorusso A., Perrone A., Strafella F., Dutouquet C., Torralba B., 2015, Phys. Rev. E, 92, 053103
Hinkel N.R., et al., 2016, ApJS, 226, 4
Iwao M., JayaRam K., 1971, Acta Phys. Polonica, A40, 527
Konjević N., 1999, Physics Reports, 316, 339
Konjević N., Dimitrijević M.S., Wiese W.L., 1984a, J. Phys. Chem. Ref. Data, 13, 619
Konjević N., Dimitrijević M.S., Wiese W.L., 1984b, J. Phys. Chem. Ref. Data, 13, 649
Konjević N., Ivković M., Jovićević S., 2010, Spectrochim. Acta Part B, 65, 593

Konjević N., Lesage A., Fuhr J.R., Wiese W.L., 2002, J. Phys. Chem. Ref. Data, 31, 819
Konjevic N., Roberts J.R., 1976, J. Phys. Chem. Ref. Data, 5, 209
Konjević N., Wiese W.L., 1976, J. Phys. Chem. Ref. Data, 5, 259
Konjević N., Wiese W.L., 1990, J. Phys. Chem. Ref. Data, 19, 1307
Konjević R., Konjević N., 1986, Fizika, 18, 327
Kopfermann H., Schneider E.E., 1958, Academic Press, New York
Kramida A., Ralchenko Y., Reader J., NIST ASD Team, 2018. NIST Atomic Spectra Database (ver. 5.6.1), [Online]. Available: https://physics.nist.gov/asd.
Krief M., Feigel A., Gazit D., 2016, ApJ, 824, 98
Kurucz R.L., Bell B., 1995. Smithsonian Astrophysical Observatory, Cambridge, Mass.,
Labutin T.A., Zaytsev S.M., Popov A.M., Seliverstova I.V., Bozhenko S.E., Zorov N.B., 2013, Spectrochim. Acta Part B, 87, 57
Lesage A., 2009, New Astron. Rev., 52, 471
Man B.Y., Dong Q.L., Liu A.H., Wei X.Q., Zhang Q.G., He J.L., Wang X.T., 2004, J. Optics A, 6, 17
Mishenina T.V., Gorbaneva T.I., Basak N.Y., Soubiran C., Kovtyukh V.V., 2011, Astron. Rep., 55, 689
Ovechkin G.V., Sandrigailo L.E., 1969, J. Appl. Spectrosc., 10, 372
Pichler G., 1972, Fizika, 4, 235
Popov A.M., Akhmetzhanov T.F., Labutin T.A., Zaytsev S.M., Zorov N.B., Chekalin N.V., 2016, Spectrochim. Acta Part B, 125, 43
Popov A.M., Labutin T.A., Zaytsev S.M., Zorov N.B., 2017, Opt. Spectrosc., 123, 521
Radzig A.A., Smirnov B.M., 1985, Springer-Verlag, Berlin
Sahal-Bréchot S., Dimitrijević M.S., Ben Nessib N., 2017a, Open Astron., 20, 523
Sahal-Bréchot S., Dimitrijević M.S., Moreau N., 2017b. Observatory of Paris, LERMA, and Astronomical Observatory of Belgrade,
Shi J.R., Gehren T., Zeng J.L., Mashonkina L., Zhao G., 2014, ApJ, 782, 80
Short C.I., Jason H.T.B., Lindsey M.B., 2018, ApJ, 854, 82
Skočić M., Burger M., Nikolić Z., Bukvić S., Djeniže S., 2013, J. Phys. B, 46, 185701
Song K., Cha H., Lee J., Lee Y.-I., 1999, Microchem. J., 63, 53
Tagirov R.V., Shapiro A.I., Schmutz W., 2017, A&A, 603, A27
Tenenbaum J., Smilanski I., Gabay S., Levin L.A., Erez G., Lavi S., 1980, Optics Commun., 32, 473
Tonejc A.M., Vujnović V., 1969, 9th Int. Conf. Phenomen. Ionized Gases, Contributed papers. Bucharest, p. 520
van Regemorter H., 1962, ApJ, 136, 906
Yan H.L., Shi J.R., Nissen P.E., Zhao G., 2016, A&A, 585, A102
Yan H.L., Shi J.R., Zhao G., 2015, ApJ, 802, 36
Zaytsev S.M., Popov A.M., Labutin, 2019, Spectrochim. Acta Part B, 158, 105632
Zaytsev S.M., Popov A.M., Zorov N.B., Labutin T.A., 2014, J. Instrum., 9, P06010
Zhao G., et al., 2016, ApJ, 833, 225
Zmerli B., Ben Nessib N., Dimitrijević M.S., Sahal-Bréchot S., 2010a, Mem. S.A.It. Suppl., 15, 152
Zmerli B., Nessib N.B., Dimitrijević M.S., Sahal-Bréchot S., 2010b, Phys. Scripta, 82, 055301